\documentclass[letterpaper, 10 pt, journal]{ieeetran}  % Comment this line out
\IEEEoverridecommandlockouts                              % This command is only
\usepackage[utf8]{inputenc}
\usepackage[T1]{fontenc}
\usepackage{graphicx}
\usepackage{float}
\usepackage{url}
\usepackage{siunitx}
\usepackage{subcaption}
\usepackage{xcolor}
\usepackage{comment}
\usepackage{hhline}
\usepackage{soul}
\title{\LARGE \bf
An FPGA-based System for Generalised Electron Devices Testing
}

\author{Patrick Foster$^*$, Jinqi Huang$^*$, Alex Serb$^{*\ddagger}$, Spyros Stathopoulos$^*$, Christos Papavassiliou$^{\dagger\ddagger}$, and Themis Prodromakis$^{*\ddagger}$\\
Email: \{p.foster, j.huang, a.serb, s.stathopoulos, t.prodromakis\}@soton.ac.uk c.papavas@imperial.ac.uk\\
$^*$Centre for Electronics Frontiers, Zepler Institiute, University of Southampton\\
$^\dagger$Department of Electrical and Electronic Engineering, Imperial College London
$^\ddagger$ArC Instruments
}

\begin{document}

\maketitle
\thispagestyle{empty}
\pagestyle{empty}
\sisetup{detect-weight = true}

%%%%%%%%%%%%%%%%%%%%%%%%%%%%%%%%%%%%%%%%%%%%%%%%%%%%%%%%%%%%%%%%%%%%%%%%%%%%%%%%
\begin{abstract}

Electronic systems are becoming more and more ubiquitous as our world digitises. Simultaneously, even basic components are experiencing a wave of improvements with new transistors, memristors, voltage/current references, data converters, etc, being designed every year by hundreds of R\&D groups world-wide. To date, the workhorse for testing all these designs has been a suite of lab instruments including oscilloscopes and signal generators, to mention the most popular. However, as components become more complex and pin numbers soar, the need for more parallel and versatile testing tools also becomes more pressing. In this work, we describe and benchmark an FPGA system developed that addresses this need. This general purpose testing system features a 64-channel source-meter unit (SMU), and 2x banks of 32 digital pins for digital I/O. We demonstrate that this bench-top system can obtain $\SI{170}{\pico\ampere}$ current noise floor, $\SI{40}{\nano\second}$ pulse delivery at $\pm\SI{13.5}{\volt}$ and $\SI{12}{\milli\ampere}$ maximum current drive/channel. We then showcase the instrument's use in performing a selection of three characteristic measurement tasks: a) current-voltage (IV) characterisation of a diode and a transistor, b) fully parallel read-out of a memristor crossbar array and c) an integral non-linearity (INL) test on a DAC. This work  introduces a down-scaled electronics laboratory packaged in a single instrument which provides a shift towards more affordable, reliable, compact and multi-functional instrumentation for emerging electronic technologies.

%Memristor crossbar arrays offer a novel new approach for designing high density non-volatile memory; however, precise measurement of resistive crossbar elements requires parallel current sensing capability not found in existing instruments. To provide this capability, we have designed and built an FPGA-based crossbar control instrument with independent per-channel biasing and measuring. In this paper, we cover the architecture of this new instrument, its operation and interface, and the results of testing conducted on the instruments current sensing and pulse driver circuitry.

\end{abstract}


\begin{figure*}
    \begin{subfigure}{0.05\linewidth}
        \hspace{\linewidth}
    \end{subfigure}
    \begin{subfigure}[b]{0.39\linewidth}
        \centering
        \includegraphics[width=\linewidth]{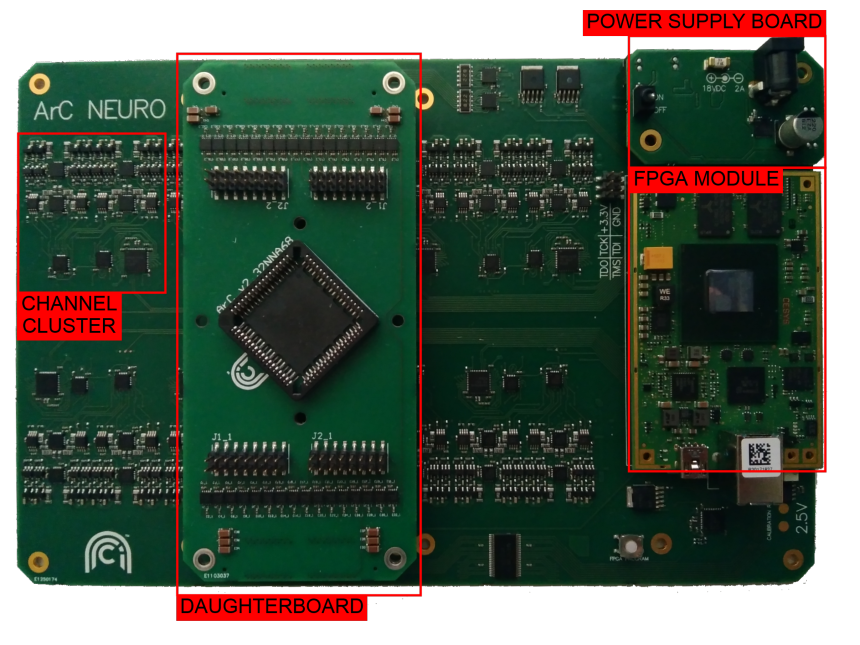}
        \caption{}
        \label{fig:arcpic}
    \end{subfigure}
    \begin{subfigure}{0.08\linewidth}
        \hspace{\linewidth}
    \end{subfigure}
    \begin{subfigure}[b]{0.39\linewidth}
        \centering
        \includegraphics[width=\linewidth]{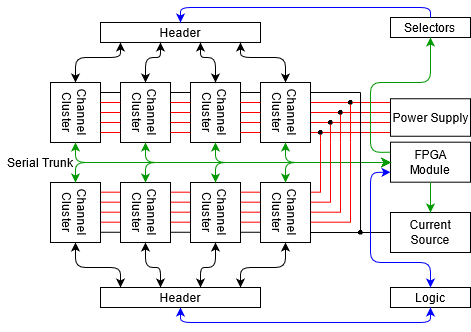}
        \caption{}
        \label{fig:brdSch}
    \end{subfigure}
    \caption{Overview of instrument. a) Picture of fully assembled system PCBs, including base board, device-under-test interfacing daughterboard, FPGA dev board and power supply board. b) High-level block diagram of system architecture illustrating parallelism and modularity of the system. Analogue connections are shown in black, serial connections are shown in green, parallel connections are shown in blue, and power supply connections are shown in red.}
\end{figure*}

%%%%%%%%%%%%%%%%%%%%%%%%%%%%%%%%%%%%%%%%%%%%%%%%%%%%%%%%%%%%%%%%%%%%%%%%%%%%%%%%
\section{Introduction}

Progress of electronic technologies has relied on a solid foundation of instrumentation tools ranging from single components, such as instrumentation amplifiers \cite{Input1998}, and high-end data converters \cite{Feedback2017}, to small-size printed circuit board (PCB) instruments for generalised parameter measurements \cite{2009char, pc-based}, benchtop instruments such as oscilloscopes and signal generators. These instruments have defined both the limits of what can be measured and tested, and play a significant role in determining the productivity of laboratories around the world. In fact, it is particularly the latter that has led to the development of specialist instrumentation such as lock-in amplifiers \cite{lock-in} and spectrum analysers \cite{Oscilloscopes}.

Over time,  both the variety and complexity of circuits being developed and requiring testing is increasing. As an example let us consider the story of instrumentation for the emerging memory devices (including memristors) community \cite{Strukov08}. These devices act as electrically tuneable resistors and hence require analogue instrumentation for their characterisation with typical tests being current-voltage sweeps, incremental-step pulse programming; see \cite{Stathopoulos2019a} for a characterisation methodology. Moreover, Resistive Random Access Memory (RRAM) memristive devices are very frequently used as crossbar arrays for performing dot products \cite{Serb2017HardwarelevelBI}. This need has led to the development of lightweight instrumentation emphasising parallelism and speed of data acquisition over raw accuracy \cite{Berdan2015b, Xing2016, Knowm}. This, in turn, has implied significant circuit design effort to mitigate effects related to sneak paths \cite{Linn2010}, which were shown to lead to potentially catastrophically undermine read-out accuracy via a variety of imperfection mechanisms \cite{Serb2015, Chen2013}. Nonetheless,  these array-level instruments were soon superseded by increasing complexity in RRAM crossbar arrays with the popularisation of the so-called `1T1R' approach \cite{Sivan2019}, where each RRAM device is paired with a `selector transistor', thus now requiring a new set of control terminals for the gates of the transistors (as shown later in Fig. \ref{fig:CBs}). In parallel, advances in RRAM technology have led to memristor cells capable of ever finer gradations of their resistive states \cite{Stathopoulos2017}, which has been pushing the accuracy requirements of instrumentation upwards.

The exemplar story of RRAM instrumentation illustrates the trend towards higher `device under test' and `circuit under test' complexity, with numerous other examples easy to draw from precision amplifiers with 20 pins \cite{opa3s328} to multi-channel switches and data converters \cite{Feedback2017}, micro-controllers \cite{Semiconductors2020} etc. In response to this trend, several designs have been dedicated to the testing systems for general electronic devices or specific devices such as RRAM. \cite{Wust2017} developed a field programmable gate array (FPGA) based memristor prototyping environment, but with a maximum theoretical resolution of $\SI{740}{\pico\ampere}$, this system cannot deliver more detailed characterisation tasks. \cite{Berdan2015b} implemented a microcontroller-based advance testing system for memristor devices, but the parallelism is limited. \cite{Wang2019} presented a high-speed driving system for phase change memory devices, with the narrowest pulse width of 500ns. However, this work only has a driver side. Other works such as \cite{Emmanuelle2016} applied commercially available device analysers, which have limited channel numbers as well as parallelism.
In continuation of our previous work in the field of RRAM instrumentation, we have developed a new instrument with the purpose of being highly parallel, competitively accurate to heavier bench-top instruments, easily transportable, and flexible enough to test circuits with up to a maximum of 128 pins with an array of analogue and digital source and metering capabilities.

In this paper, we present the scientific contributions resulting from the development of this new instrument, namely: i) the design and implementation of a general purpose, 64-channel fully-parallel analogue source-metre unit (SMU) with specialist circuitry introduced to allow (a) current-mode biasing and (b) high-speed pulsing capability (tackled in section \ref{sec:sysimp}) and ii) the benchmarking of the SMU's performance in terms of accuracy, noise floor and pulsing characteristics (section \ref{sec:res}). Furthermore, we illustrate how the instrument can be used flexibly via presenting three practical examples: characterising a transistor, interfacing a RRAM crossbar array and testing the differential non-linearity (DNL) of a data converter (in section \ref{sec:apps}) and conclude the paper (section \ref{sec:disc}) by discussing the opportunities arising.

\section{SYSTEM IMPLEMENTATION}\label{sec:sysimp}

The system we have developed is shown in Fig. \ref{fig:brdSch}. It comprises a 64-channel, fully parallel SMU array and 2x banks of 32 digital pins. The instrument also features a shared current source.   The entire system is coordinated by an FPGA EFM-03 development board with Xilinx XC7A200T-2FBG676I chip and is controlled by a PC via a python-based interface. The system has been engineered to provide high-speed, parallel testing at high-levels of accuracy. The assembled instrument is shown in Fig. \ref{fig:arcpic}, with the standard interfacing daughter-board (for connecting to PLCC68 packages). The power supply daughter-board and FPGA development board can also be seen.

\begin{figure}[H]
    \centering
    \includegraphics[width=0.95\linewidth]{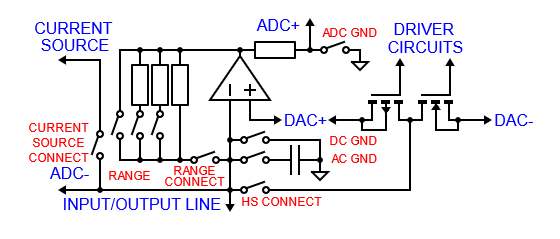}
    \caption{Schematic of the channel architecture. Significant wires are labelled in blue. Analogue switches are labelled in red.}
    \label{fig:channelsch}
\end{figure}
\subsection{Subsystem overview}\label{subsec:sysover}

The main subsystem of the board is the SMU channel. It consists of: a) a programmable gain trans-impedance amplifier (TIA), b) an independent pulse generator used for high-speed pulsing and c) a switch which allows the channel to access the current source, as shown in Fig. \ref{fig:channelsch}. Data converter terminals are connected as shown in Fig. \ref{fig:channelsch} to provide biasing with digital to analogue converters (DACs), allowing the channel to act as a tuneable source, or to read voltages with differential analogue to digital converters (ADCs) at selected nodes, allowing the module to act as a measurement unit.

The TIA structure is designed to act as either a source or a meter for the Input/Output (I/O) line node. In voltage source-mode, the feedback by-pass switch RANGE and feedback connection switch RANGE CONNECT are both closed and every other switch is open. The DAC sets the desired sourcing voltage and ADC+ measures what is actually being applied. As a special case, to ground the I/O line all switches remain open except DC GND, which creates the shunt to real GND. In current meter mode, on the other hand, RANGE CONNECT and an appropriate combination of feedback resistor bank switches RANGE are connected. Any current from the I/O line is converted into a voltage by TIA action and measured by ADC+. For direct voltage metering, ADC GND is closed and a direct measurement is taken at ADC-. A shunt-to-ground capacitor can be connected at any time to protect against sudden changes in voltage, or even just erode the noise bandwidth.

The amplifier selected for this design was chosen for its low quiescent current and input bias current, but this comes at the cost of frequency response, with a gain-bandwidth product of only $\SI{2.5}{\mega\hertz}$. The analogue switches were chosen for a balance of low on resistance and charge injection, with $\SI{9.5}{\ohm}$ and $\SI{4}{\pico\coulomb}$ respectively.

%\textcolor{orange}{The TIA has selectable ranges, and is referenced against a DAC channel, allowing the amplifier to set its input node to any voltage. This can be used for high current biasing, by using the short circuit switch in the amplifier ranging array. The differential ADC is connected to the input (through a low current bias op-amp buffer) and the output, guaranteeing that the measured value is representative of the current into the input node, regardless of the voltage at the input. The output of the amplifier can be grounded, changing the channel into a voltage sensor. The high resolution of the ADC and the low input current bias permit sub $\SI{}{\nano\ampere}$ measurements without requiring large arrays of ranging resistors.}

The high-speed pulse driver is implemented with a complementary MOSFET pair which can drive the output line to the voltage of either of two DAC channels. This connectivity allows for variable pulse amplitude and enables high-speed by keeping the path between charge supplier and I/O line very simple and low impedance. Bi-phasic pulses can be constructed across 2-terminal devices by having two channels swing between $V_+>0$ and $0$ and $V_-<0$ and $0$ respectively.

The switch CURRENT SOURCE CONNECT connects the I/O line to the shared current source, to permit current biasing. Current biasing can also be achieved through the TIA by using successive approximation, if parallel operation is required, but precise current control requires a dedicated circuit which is too large to be included in the channel. As a result of sharing, a more complex dedicated current source could be designed to source or sink sub-$\SI{}{\nano\ampere}$ currents, permitting current biasing of $\SI{}{\giga\ohm}$ scale resistive devices. The current source circuit also contains a precision voltage reference, which can be connected to any channel to calibrate the ADC. At a higher level, the individual SMU channels are grouped into clusters of eight. This allows each cluster to share one 8 channel 18-bit ADC and one 16 channel 16-bit DAC. The analogue switch IC used in this design has an integrated serial FIFO register, allowing the switches of all the channels in a cluster to be controlled in single serial daisy-chain. The switch, ADC, and DAC serial lines from each cluster are grouped together into a bus that runs down the centre of the board, called the serial trunk. The cluster is physically arranged so that all control signals are on one side, with the measurement lines on the other and the supply rails running perpendicular on a different layer. Each cluster also shares the same control signals for the high-speed drivers. Channels in separate clusters can produce asynchronous pulses, but channels in the same cluster cannot.

The next subsystem is the digital pin banks. The first bank of 32 channels (the `selector' bank) is an output-only set that is intended to drive transistor gates. This was developed to address the needs of selector transistors in RRAM arrays \cite{Jouppi2011}. As a result, the HI and LO voltages can be set arbitrarily, but they are common for the entire bank. Furthermore, both drive strength and speed are relatively low. The second bank (the `arbitrary logic bank') is a more conventional full digital I/O system, which is referenced exclusively to GND. It is intended to drive digital pins on test chips or read from them.

\subsection{Digital interface hierarchy}
\begin{figure}[H]
    \centering
    \includegraphics[width=\linewidth]{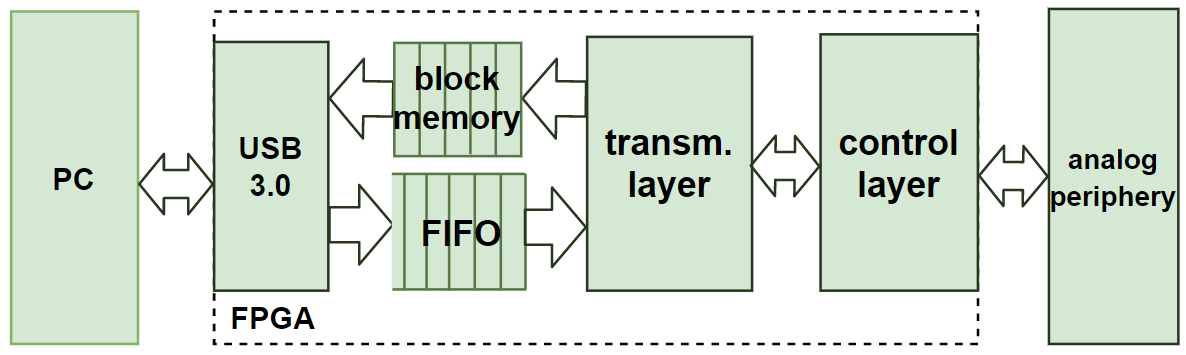}
    \caption{Digital interface hierarchy. The speed of internal data bus is 3.2Gbps.}
    \label{fig:digital}
\end{figure}

Fig. \ref{fig:digital} illustrates the concept diagram of the digital interface, which bridges the gap between the PC-level software and the analogue circuitry of the PCB board. The basic structure of the digital interface contains a USB 3.0 software core, a first-in-first-out (FIFO) buffer, block memory, a transmission layer and a control layer. The instruction set has been designed for translating a relatively small set of high-level operations into "board language". These are: select channels, emit pulse, read from channel(s) as well as set current (for the shared current source) and a few more specialised commands. In hardware, this translates to configuring the high-speed pulse drives, DACs, ADCs, switches and digital pins. All advanced functions can be performed through a combination of the basis set of commands. The transmission layer performs the translation from PC-level instructions to PCB-level and the control layer executes the latter.

As an example, a basic write operation needs commands to configure the high-speed pulse driver and SMU channel switches (see Fig. \ref{fig:channelsch}). Information such as voltage pulse amplitude, pulse width and target devices will be processed and converted on the PC. Then, the FPGA will receive the commands through USB3.0, configure the target channel and then trigger the pulse. Information flows in the opposite direction in a basic read operation. Commands for DACs and ADCs are sent to configure the bias voltage and start voltage readout in the selected channels. The measurement results are temporarily stored in the on-chip memory of the FPGA waiting for the PC to be ready to process it. To match the PC-side and FPGA-side speeds of transmission and processing, a FIFO buffers the PC-to-FPGA downlink and a block memory buffers the uplink. The FIFO can currently fit just one instruction package, but will eventually be upgraded to 32+ instructions.

All IPs inside the FPGA are linked through an Advanced Advanced eXtensible Interface (AXI). AXI is a universal high-speed high-performance interface, typically used in microcontroller systems \cite{Limited2021}. The burst-based property of AXI and 100MHz FPGA system clock allows internal data transmission rates of up to 3.2Gbps . The third-party USB3.0 IP\cite{cesys} we used also generated 100MHz clock for USB controller chip CYUSB3014 \cite{Pattnayak}, giving the same maximum 3.2Gbps data rate for communication via usb. 

\section{EXPERIMENTAL RESULTS}\label{sec:res}

Benchmarking the instrument involved performing a set of experiments to determine the noise floor of voltage and current read operations, the read-out accuracy of test resistances, the pulse characteristics obtained at the when using the write functions of the system and some basic data on the functionality of the digital terminals.

\subsection{Noise floor}

\begin{figure}[t]
    \centering
    \begin{subfigure}[b]{0.49\linewidth}
        \centering
        \includegraphics[width=\linewidth]{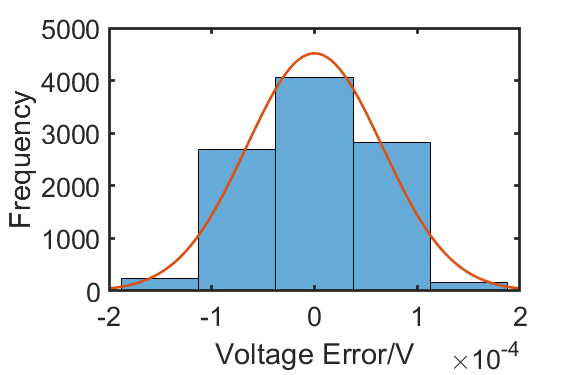}
        \caption{}
        \label{fig:VOLTnoise}
    \end{subfigure}
    \begin{subfigure}[b]{0.49\linewidth}
        \centering
        \includegraphics[width=\linewidth]{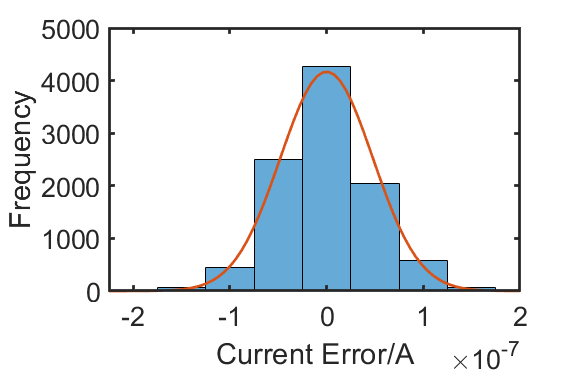}
        \caption{}
        \label{fig:820noise}
    \end{subfigure}
    \begin{subfigure}[b]{0.49\linewidth}
        \centering
        \includegraphics[width=\linewidth]{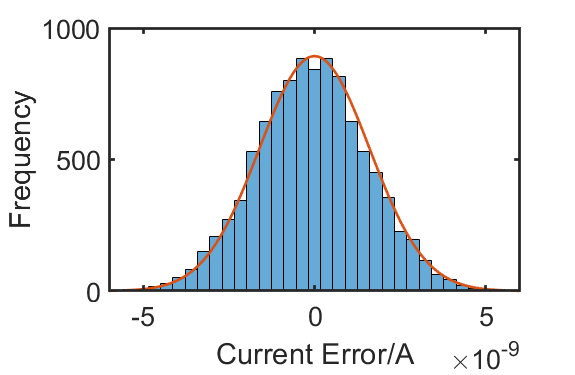}
        \caption{}
        \label{fig:110noise}
    \end{subfigure}
    \begin{subfigure}[b]{0.49\linewidth}
        \centering
        \includegraphics[width=\linewidth]{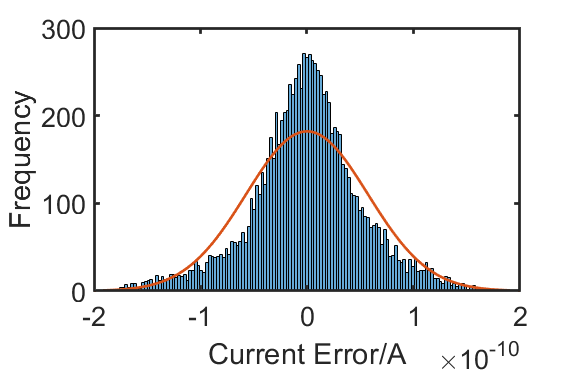}
        \caption{}
        \label{fig:15noise}
    \end{subfigure}
    \caption{Histograms showing noise characteristics of the various modes of measurement. All histograms have one bin per ADC code with widths of $\SI{78.1}{\micro\volt}$, $\SI{47.6}{\nano\ampere}$, $\SI{355}{\pico\ampere}$, and $\SI{2.60}{\pico\ampere}$ respectively. a) 10k point histogram of a read-out voltage error test (V=GND), overlaid with Gaussian distribution estimate. We obtain $\sigma=\SI{65}{\mu\volt}$. b-d) 10k point histograms of current read-out tests, overlaid with Gaussian distribution estimates. b: $\SI{820}{\ohm}$ TIA range yields $\sigma=\SI{48}{\nano\ampere}$. c: the $\SI{110}{\kilo\ohm}$ TIA range yields $\sigma=\SI{1.6}{\nano\ampere}$. d: the $\SI{15}{\mega\ohm}$ TIA range yields $\sigma=\SI{57}{\pico\ampere}$.}
    \label{fig:noise}
\end{figure}

To assess the noise floor of voltage readings, we grounded a channel and collected 10k voltage readings as shown in Fig. \ref{fig:VOLTnoise}. The voltage readings mostly spanned across three consecutive ADC codes. Using a Gaussian noise model we estimated standard deviation (s.d.) of $\SI{65}{\mu\volt}$, although this may not be accurate due to the variance being of similar scale to the quantisation error.

To assess the noise floor of current readings, we configured a channel as a TIA with a reference of $\SI{-0.5}{\volt}$. We then connected different resistors between the input node and ground to produce a bias current that forces the channel to automatically select a specific range, and then collected 10k readings in each range, as shown in Fig. \ref{fig:noise}. The uncertainty in the instrument's readings was thus obtained. For the $\SI{820}{\ohm}$ range, we connected a $\SI{2.2}{\kilo\ohm}$ resistor. As with the voltage readings, in this range the results mostly spanned across just three consecutive ADC codes (Fig. \ref{fig:820noise}). This suggests to us that the noise in this range is dominated by the ADC noise and quantisation error. Using a Gaussian noise model we estimated s.d. of $\SI{48}{\nano\ampere}$. The test was repeated with a $\SI{16.4}{\kilo\ohm}$ resistor, targeting the $\SI{110}{\kilo\ohm}$ TIA gain range (Fig. \ref{fig:110noise}). The distribution was approximately Gaussian, with an s.d. of $\SI{1.6}{\nano\ampere}$, or roughly 5 LSB. To test the $\SI{15}{\mega\ohm}$ TIA range, we left the TIA input open circuit and obtained s.d. of $\sigma=\SI{57}{\pico\ampere}$, or roughly 22 LSB. The error distribution in this range did not display the Gaussian distribution obtained in tests of other ranges. Experimentation showed that the extended tail of the distribution was a result of mains interference: During each test, the wires used to connect resistors for preceding tests was left in place; removing these -thereby reducing the length of floating input line- resulted in reduced uncertainty. The input lines of the channel act as an antenna, collecting energy emitted from nearby mains wiring. All results presented here represent the test wires removed to minimise antenna effects. This could likely be eliminated by operating the instrument inside an anechoic chamber, adding load capacitance or other good measurement techniques applied independent of the instrument.

\subsection{Read-out accuracy}

\begin{figure}[t]
    \centering
    \includegraphics[width=\linewidth]{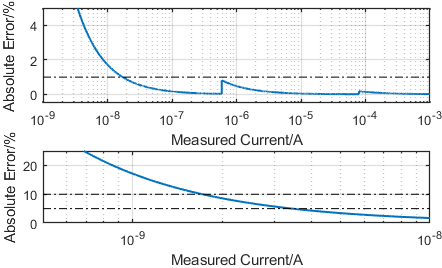}
    \caption{Graph showing predicted absolute error based on $3\sigma$ current noise error.}
    \label{fig:currerr}
\end{figure}

We calculated the `reasonable worst-case' proportional current reading error across the designed operating range of the instrument by assuming a measurement error of $3\sigma$ (Fig. \ref{fig:currerr}). Current measurements of more than $\SI{16}{\nano\ampere}$ can be made with 1\% accuracy, at a sampling rate of $\SI{833}{\hertz}$. Measurements above $\SI{3.4}{\nano\ampere}$ and $\SI{1.7}{\nano\ampere}$ can be made with 5 and 10\% accuracy respectively. The calculation suggests that, at a bias voltage of $\SI{0.5}{\volt}$, we can read resistance of devices up to $\SI{100}{\mega\ohm}$ before precision starts to degrade. With further averaging, it may be possible to push the maximum resistance up to $\approx\SI{1}{\giga\ohm}$, but diminishing returns will impose practical limits. The effect of changing ranging resistors is clearly visible in the figure as step discontinuities in the error magnitude.

\subsection{Pulse characteristics} 

\begin{figure*}[ht]
    \centering
    \begin{subfigure}[b]{0.245\textwidth}
        \centering
        \includegraphics[width=\textwidth]{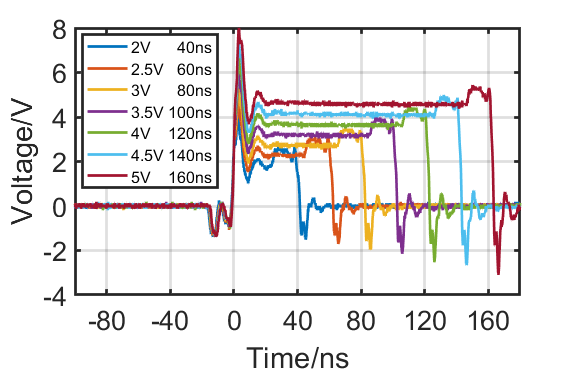}
        \caption{}
    %    \caption{+VE pulses starting at $\SI{0}{\volt}$}
        \label{fig:pls1}
    \end{subfigure}
    \begin{subfigure}[b]{0.245\textwidth}
        \centering
        \includegraphics[width=\textwidth]{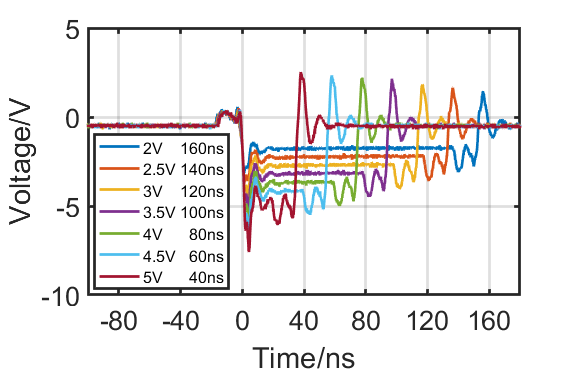}
        \caption{}
    %    \caption{-VE pulses starting at $\SI{-0.5}{\volt}$}
        \label{fig:pls2}
    \end{subfigure}
    \begin{subfigure}[b]{0.245\textwidth}
        \centering
        \includegraphics[width=\textwidth]{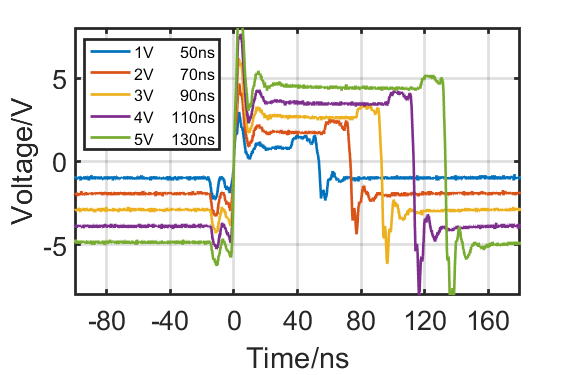}
        \caption{}
    %    \caption{+VE pulses symmetrical around $\SI{0}{\volt}$}
        \label{fig:pls3}
    \end{subfigure}
    \begin{subfigure}[b]{0.245\textwidth}
        \centering
        \includegraphics[width=\textwidth]{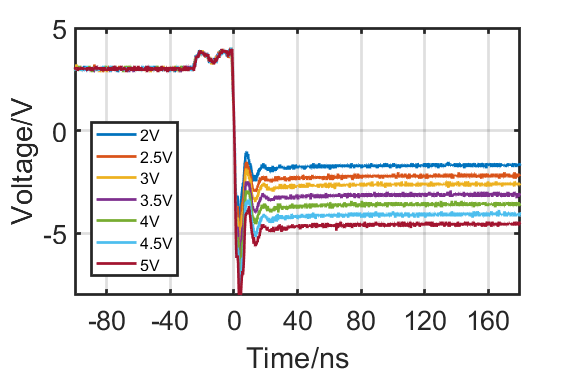}
        \caption{}
    %    \caption{Continuous pulses starting at $\SI{3}{\volt}$}
        \label{fig:pls4}
    \end{subfigure}
    \caption{Oscilloscope captures of a variety of pulses produced with the high speed pulse generator. (a) +VE pulses starting at $\SI{0}{\volt}$. (b) -VE pulses starting at $\SI{-0.5}{\volt}$. (c) +VE pulses symmetrical around $\SI{0}{\volt}$. (d) Continuous pulses starting at $\SI{3}{\volt}$}
    \label{fig:plsGraphs}
\end{figure*}

Here, we tested the quality of short duration pulses produced by the high speed drivers, as well as the delay mismatch between channels. We commanded a range of pulses with varying high and low values, in increments of $\SI{10}{\nano\second}$ between the minimum pulse width ($\SI{40}{\nano\second}$) and $\SI{160}{\nano\second}$. Although the instrument is capable of producing pulses with high and low states anywhere within the range of the DACs at $\pm\SI{13.5}{\volt}$, we were only able to test pulses between $\pm\SI{5}{\volt}$ due to limits of the high speed probes that were available. Repetition rates above $\SI{1}{\mega\hertz}$ were found to cause significant heating in the driver circuits during prolonged testing, but shorter pulse trains with a repetition rate of up to $\SI{12.5}{\mega\hertz}$ should be possible. The rise and fall times were all comparable, at $2-\SI{4}{\nano\second}$ (Fig. \ref{fig:plsGraphs}). We observed a maximum mismatch of $\SI{1.5}{\nano\second}$ between channels. This is small enough to enable differential write operations (for example the biphasic pulses described in section \ref{subsec:sysover}).

\subsection{Digital terminals}

The instrument has two banks of digital channels (Table \ref{tab:digBanks}): a 'selector' bank of 32 serially addressed digital outputs and an 'arbitrary level logic' bank of 32 IO pins.

The `selector' bank supports HI and LO voltages anywhere within the full $\pm\SI{13.5}{\volt}$ range at a guaranteed minimum resolution of $\SI{600}{\micro\volt}$. Rise times are determined by an $~\SI{100}{\nano\second}$ switch closing time plus the RC defined by the on-resistance of the switch ($\SI{9.5}{\ohm}$). Fall times are determined by the RC of a pull-down circuit with $R_{PD} = \SI{8.2}{\kilo\ohm}$. The circuit is configured in such way that the user can set the nominal HI voltage to be lower than LO, thereby swapping the roles of the switch and the pull-down/up resistor. This can be used, for example when a very fast fall time is required. The minimum pulse length on any pin is approx. $\SI{1.3}{\micro\second}$. This is limited by the time required to write to the serial registers that control the selector states.

The `arbitrary level logic' bank is a more conventional array of bidirectional level shifter ICs, with a selectable HI level of between 1.8-5.5V, at a resolution of $\SI{120}{\micro\volt}$. This bank is operated in parallel directly from the FPGA IO pins and as a result can operate at much higher frequency than the selector bank. The level shifters have typical rise and fall time of between $\SI{1.3}{\nano\second}$ and $\SI{4}{\nano\second}$, depending on the voltage level set. The typical propagation delay is also dependent on the selected voltage level and is typically below $\SI{8}{\nano\second}$, except at very low voltage levels, where the delay in output configuration may be as high as $\SI{20}{\nano\second}$.

\begin{table}[h]
    \centering
    \resizebox{0.6\linewidth}{!}{    
        \begin{tabular}{|c||c|c|}
        \hline
         & Selectors & Arbitrary logic\\
        \hhline{|=||=|=|}
        No. of channels & 32 & 32\\
        \hline     
        High value range & $\pm\SI{13.5}{\volt}$ & $\SI{1.8}{\volt}-\SI{5.5}{\volt}$\\
        \hline     
        Low value range & $\pm\SI{13.5}{\volt}$ & $\SI{0}{\volt}$\\
        \hline 
        Direction & Output & Input/Output\\
        \hline
        \end{tabular}
    }
    \caption{Selectors and arbitrary logic specifications.}
    \label{tab:digBanks}
\end{table}

\section{Application examples}\label{sec:apps}

In order to illustrate the general and versatile nature of the developed instrument we have performed a set of three example tasks as shown below. First, a classical component characterisation routine was ran on a resistor, a diode, and a transistor. Second, a set of read-out operations were conducted on a crossbar array. Third, the I/O characteristics and DNL of a DAC IC were measured. This set of tasks covers a broad range of communities ranging from device development and emerging technologies to more traditional circuit design and component testing. All single component tests were conducted using a ZIF socket daughterboard as shown in Fig. \ref{fig:dacPic}.

\subsection{Diode and transistor characterisation}

\begin{figure}[h]
    \centering
    \begin{subfigure}[b]{0.49\linewidth}
        \centering
        \includegraphics[width=\linewidth]{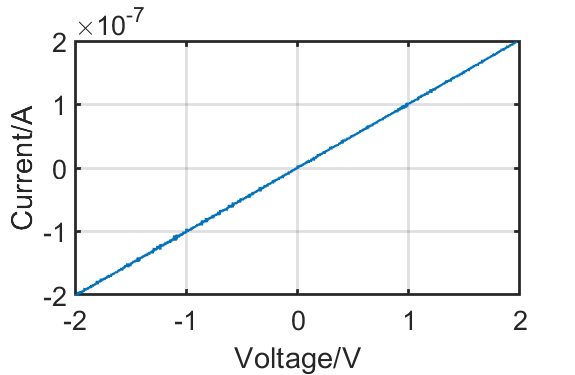}
        \caption{}
        \label{fig:resIV}
    \end{subfigure}
    \begin{subfigure}[b]{0.49\linewidth}
        \centering
        \includegraphics[width=\linewidth]{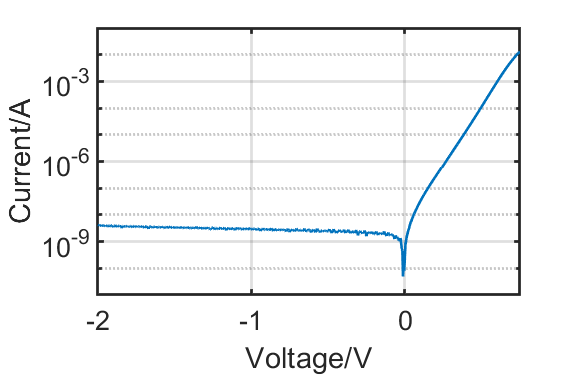}
        \caption{}
        \label{fig:diodeIV}
    \end{subfigure}
    \begin{subfigure}[b]{0.49\linewidth}
        \centering
        \includegraphics[width=\linewidth]{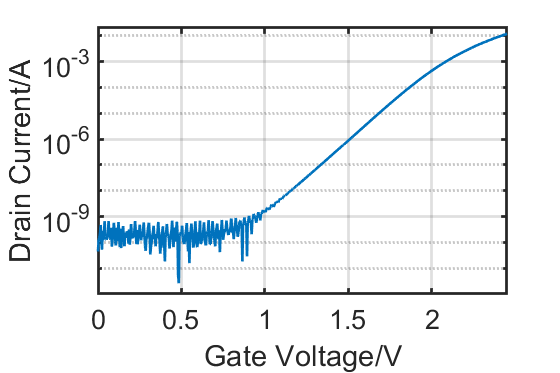}
        \caption{}
        \label{fig:fetGT}
    \end{subfigure}
    \begin{subfigure}[b]{0.49\linewidth}
        \centering
        \includegraphics[width=\linewidth]{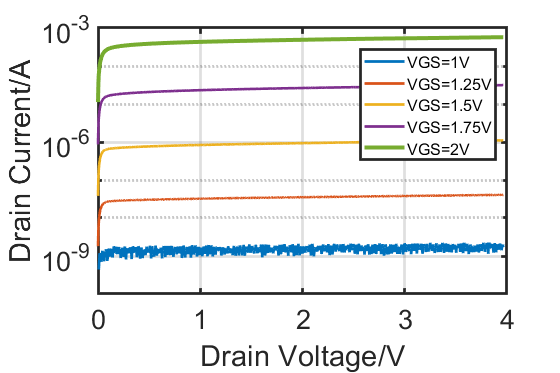}
        \caption{}
        \label{fig:fetDR}
    \end{subfigure}
    \caption{IV characteristics of a small selection of components. a) IV sweep of a $\SI{10}{\mega\ohm}$ resistor. b) IV sweep of a 1N4148 diode, from $\SI{-2}{\volt}$ to $\SI{0.75}{\volt}$. c) Gate terminal and d) drain terminal sweeps of a 2N7000 nFET.}
\end{figure}

First, we connected a $\SI{10}{\mega\ohm}$ resistor between two SMU channels and demonstrated IV sweep capability. One channel was configured to drive an arbitrary voltage, and the other was configured to measure current. An IV sweep between $\pm\SI{2}{\volt}$ was conducted, with steps of $\SI{4}{\milli\volt}$. Results are shown in Fig. \ref{fig:resIV}. The same test was then conducted with a 1N4148 small-signal diode (Fig. \ref{fig:diodeIV}). We observe that all results are above the noise floor, even in the reverse bias range. In the diode test, data points from $\SI{0.75}{\volt}$ and up have been omitted from this figure, as the rapidly increasing forward bias current saturates the TIA and the voltage across the diode is no longer controlled.

Next, we tested a 2N7000 nFET, requiring simultaneous control of three SMU channels. First, we set its drain-source voltage $V_{DS}$ to $\SI{1}{\volt}$ and swept the gate-source voltage $V_{GS}$ between 0-4V resulting in Fig. \ref{fig:fetGT}. At below approx. 1V we hit the noise floor whilst above approx. 2.4V we hit soft compliance as with the diode beforehand. Second, we performed a set of $V_{DS}$ sweeps between 0-4V at different $V_{GS}$ levels as shown in Fig. \ref{fig:fetDR}.

\subsection{Resistive crossbar handling}

\begin{figure*}[!ht]
    \centering
    \begin{subfigure}[b]{0.31\linewidth}
        \centering
        \includegraphics[width=\textwidth]{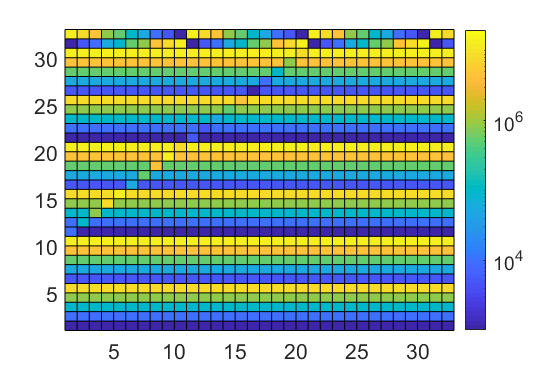}
        \caption{}
        \label{fig:array}
    \end{subfigure}
    \begin{subfigure}[b]{0.31\linewidth}
        \centering
        \includegraphics[width=\textwidth]{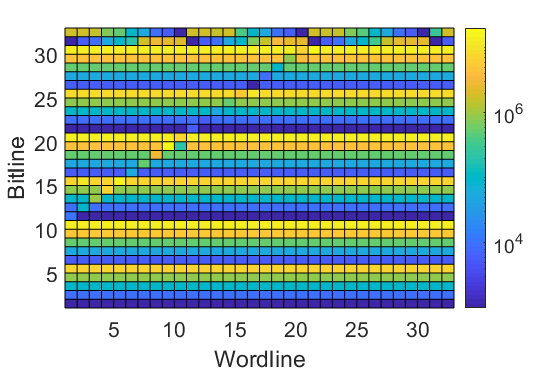}
        \caption{}
        \label{fig:wrd}
    \end{subfigure}
    \begin{subfigure}[b]{0.31\linewidth}
        \centering
        \includegraphics[width=\textwidth]{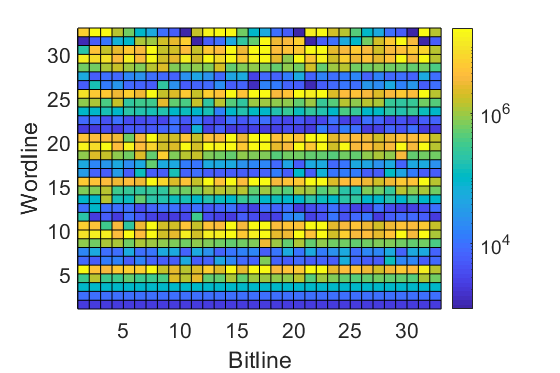}
        \caption{}
        \label{fig:bit}
    \end{subfigure}
    \begin{subfigure}[b]{0.31\linewidth}
        \centering
        \includegraphics[width=0.8\textwidth,trim={0 -4cm 0 0}]{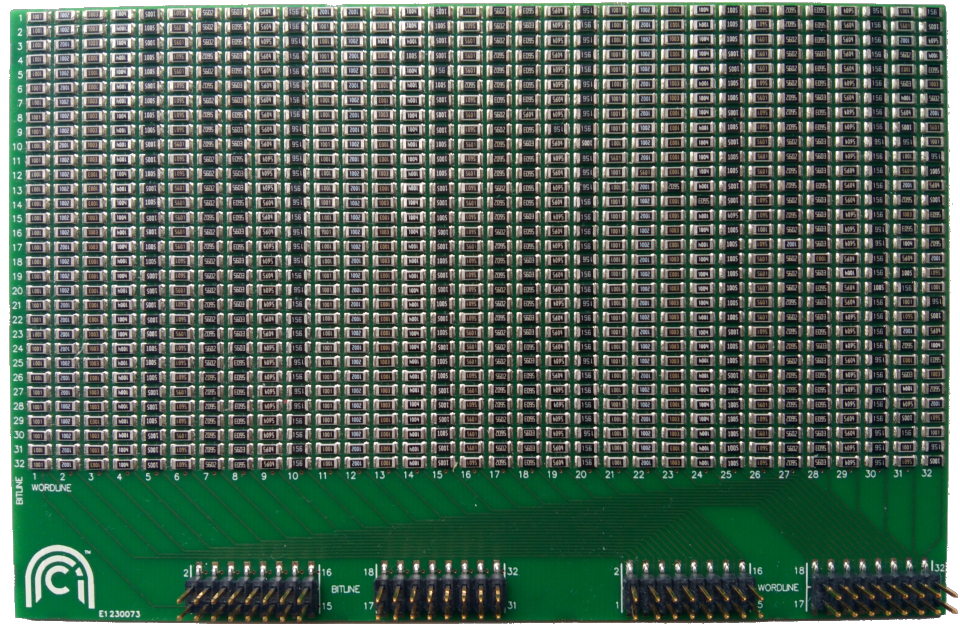}
        \caption{}
        \label{fig:resArray}
    \end{subfigure}
    \begin{subfigure}[b]{0.31\linewidth}
        \centering
        \includegraphics[width=\textwidth]{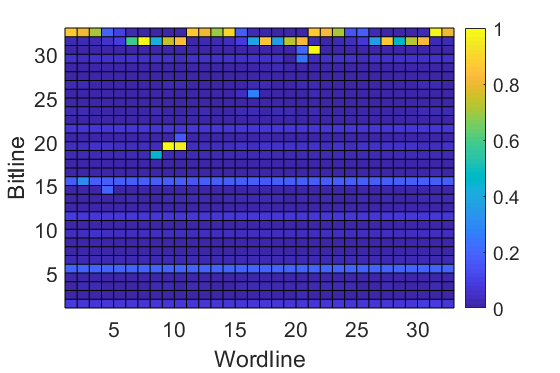}
        \caption{}
        \label{fig:wrdErr}
    \end{subfigure}
    \begin{subfigure}[b]{0.31\linewidth}
        \centering
        \includegraphics[width=\textwidth]{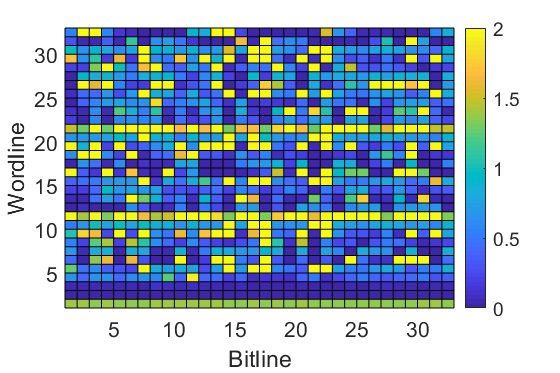}
        \caption{}
        \label{fig:bitErr}
    \end{subfigure}
    \caption{Array read operations for a 32x32 resistor array. Figure. \ref{fig:array} shows the array as designed, with resistors ranging from $\SI{1}{\kilo\ohm}$ to $\SI{15}{\mega\ohm}$. The colourbar is scaled from $\SI{1}{\kilo\ohm}$ to $\SI{20}{\mega\ohm}$ Figure. \ref{fig:wrd} shows the array as read in columns. Figure. \ref{fig:wrdErr} shows the proportional error of Figure. \ref{fig:wrd}. Figure. \ref{fig:bit} shows the array as read in rows. Figure. \ref{fig:bitErr} shows the proportional error of Figure. \ref{fig:bit}.}\label{fig:arrayMeas}
\end{figure*}

\begin{figure}
    \centering
    \begin{subfigure}[b]{0.24\linewidth}
        \centering
        \includegraphics[width=\linewidth]{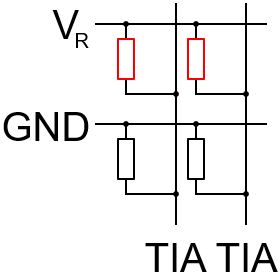}
        \caption{}
        \label{fig:read}
    \end{subfigure}
    \begin{subfigure}[b]{0.24\linewidth}
        \centering
        \includegraphics[width=\linewidth]{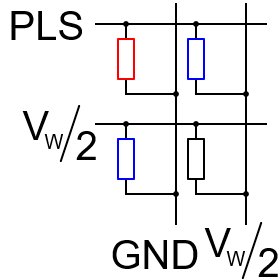}
        \caption{}
        \label{fig:write}
    \end{subfigure}
    \begin{subfigure}[b]{0.24\linewidth}
        \centering
        \includegraphics[width=\linewidth]{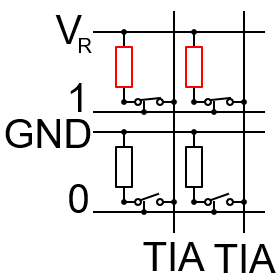}
        \caption{}
        \label{fig:selread}
    \end{subfigure}
    \begin{subfigure}[b]{0.24\linewidth}
        \centering
        \includegraphics[width=\linewidth]{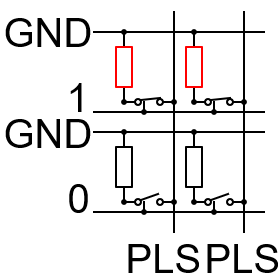}
        \caption{}
        \label{fig:selwrite}
    \end{subfigure}
    \caption{Basic read (a) and write (b) operations for selectorless crossbar arrays. (c) and (d) show the same operations for selector enabled arrays. Red, blue and black devices correspond to selected, half-selected and unselected devices.}
    \label{fig:CBs}
\end{figure}

The instrument is capable of controlling crossbar arrays and conducting parallel read and write operations. The general read and write configurations used in the RRAM community form an illustrative and instructive set of tasks for showcasing what array-level computation frequently involves. Fig.  \ref{fig:CBs} shows some examples of reading from and writing to a selectorless crossbar array (a-c), as well as interfacing an array featuring transistor-based selector devices for either reading or writing. In all cases the array can be conceptually split into the `active wordline' where bias is applied, the `active bitline', from which we may choose to measure, and the inactive word- and bitlines that need to be handled appropriately for avoiding sneak path issues. In the case of the selector-based array we also need to control the selector terminals.

In general, word- and bitlines require analogue control (both for applied voltage/current and read-out) whilst the selector terminals can be used in either modes. SMU channels can be mapped to any line requiring analogue control and enables all operating options shown in Fig. \ref{fig:CBs} as well as others (e.g. where we write by raising the active wordline to $+V_{WRITE}/2$, setting the active bitline to $-V_{WRITE}/2$ and keeping all inactive lines grounded). To write, either the high speed pulse generators or the TIA can be used. Arbitrary waveforms and slower pulses can be achieved by varying the DAC+ terminal of each channel in-operando. With 64 SMU channels the system can handle up to a 32x32 selectorless crossbar array, or 21x21 array with transistor selectors under analogue control. If the selectors can be satisfactorily controlled using the specialised, digital selector terminals, a 32x32 array with transistor selectors is supported.

%multiple read and write operations. To write, either the high speed pulse generators or the TIA can be used. In a typical write operation (Figure \ref{fig:write}), all unselected lines are biased to $V_{WRITE}/2$ and the selected bitline is grounded. The selected wordline is then pulsed to $V_{WRITE}$ using the pulse generator on that channel. Arbitrary waveforms and slower pulses can use the buffer-configured amplifier instead. The independent biasing hardware on every channel allows alternative operations, such as grounding unselected lines and setting the selected bitline to $-V_{WRITE}/2$ and pulsing wordline at $+V_{WRITE}/2$, or using differential pulses to minimise the impact on partially selected devices. \textcolor{red}{The term "differential pulses" might need explicit definition}

For this work, we chose to demonstrate a read-out on a physical 32x32 selectorless crossbar array of SMD resistors (Fig. \ref{fig:resArray}). The behaviour of an RRAM array can be approximated as a resistor for a single read voltage and this array provides known impedances from which read error can be calculated. The scheme used is seen in Fig. \ref{fig:read}: The active wordline is biased with the read-out voltage and the active bitlines are set to virtual grounds through the measurement set-up. For line-parallel read, all bitlines are active simultaneously. Multiple readings are taken and then averaged to improve precision (32 in our implementation). Naturally, line resistances and small errors in the DAC output voltages referencing the read-out SMU TIAs, etc. will all combine to introduce some errors through sneak paths. We sought to assess the extent of these imperfections in our subsequent measurements.

%\textcolor{orange}{Read operations (Figure \ref{fig:read}) bias the selected wordline at $V_{READ}$ and ground all unselected wordlines. All bitlines are configured as TIAs referenced to ground, and the current out of the entire wordline is measured in parallel. Multiple readings are taken and then averaged to improve precision (32 in our implementation). As the bitlines and unselected wordlines are held at or close to ground potential, the measured current is dominated by the current through the selected device. Errors in the DAC output voltage being used as a reference for the TIA can cause non-trivial sneak path currents in arrays where very large and very small resistance values coexist, as current can pass from one TIA to another through unselected devices \textcolor{red}{insert citation}. This issue can be eliminated through use of selector-enabled arrays \textcolor{red}{citation needed}.}

The array used 1\% resistors of $\SI{1}{\kilo\ohm}$ to $\SI{10}{\mega\ohm}$ and 5\% resistors of $\SI{15}{\mega\ohm}$; its nominal design is shown in Fig. \ref{fig:array}. To test read-out accuracy we simply performed a line-parallel read on each row and then calculated the fractional error $|(R_{meas}-R_{actual})/R_{actual}|$. Because the array is square we could use the same physical array to perform two separate tests: one on the array `as-is' and another with the array rotated by $90^o$. This allows us to illustrate the well-known issue that the value read at any point in the array depends on the states of its neighbours \cite{Serb2015}.

In the read operation we used, we found that bitline-to-bitline TIA reference mismatch degraded accuracy when trying to measure high-value resistors with low-value resistors on the same bitline. Even small differences in voltage between bitlines can cause non-trivial sneak currents to flow between them if both lines have a low resistance connection to an inactive wordline. The channel-to-channel voltage discrepancy is typically only $\SI{500}{\micro\volt}$, but if the ratio between the smallest device on a bitline and the device being read is comparable to the ratio between the read voltage and the mismatch voltage then accuracy will suffer. Our test used a read voltage of $\SI{5}{\volt}$, which gives a ratio of 10000. In a configuration where the devices on a bitline are largely of the same value (Figure. \ref{fig:wrd}) the performance is excellent, with 802 of 1024 resistors measured with less than 5\% error. Reading from the other direction (Figure. \ref{fig:bit}), the ratio between the largest and smallest devices on most bitlines is 15000. In this configuration, only 171 of 1024 resistors were measured with less than 5\% error and 758 measured with less than 100\% error. The instrument was manually calibrated for this experiment but ADC offset was not taken into account (typ. $\pm\SI{160}{\micro\volt}$). As such, the channel to channel offset voltage may be higher than expected. Automated calibration will mitigate this issue. Since the resolution of a voltage read operation is greater than the DAC resolution, it should be possible to measure the channel to channel offset and use deconvolution to obtain more accurate values, but this is beyond the scope of this paper.

\subsection{Mixed signal testing}

\begin{figure}[ht]
    \centering
    \begin{subfigure}[b]{\linewidth}
        \centering
        \includegraphics[width=0.6\linewidth]{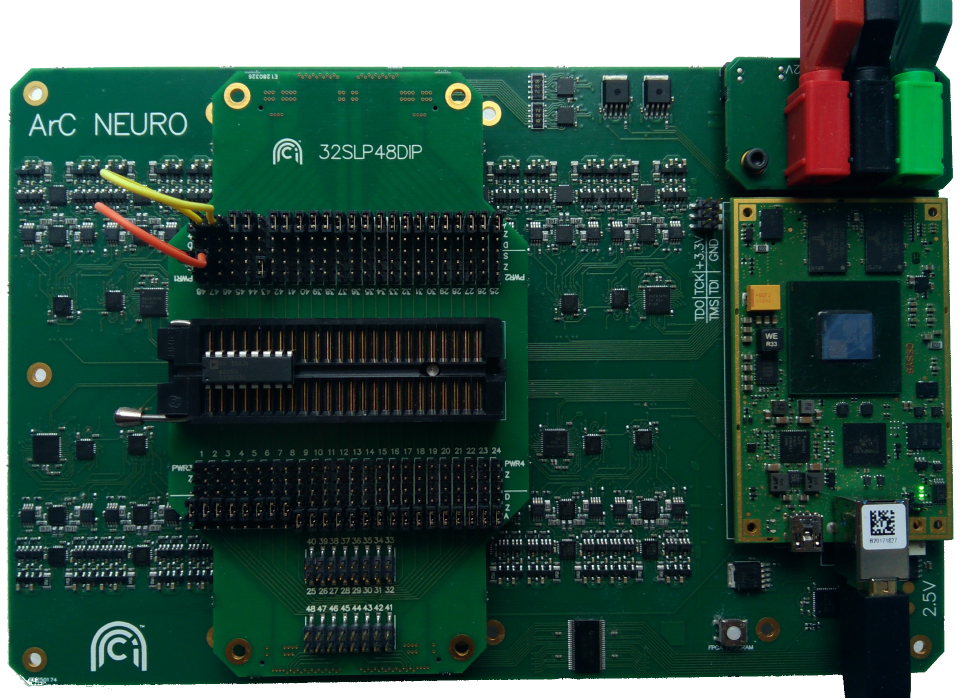}
        \caption{}
        \label{fig:dacPic}
    \end{subfigure}
    \begin{subfigure}[b]{0.49\linewidth}
        \centering
        \includegraphics[width=\linewidth]{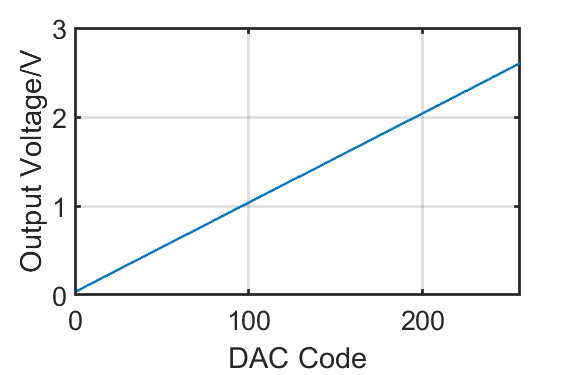}
        \caption{}
        \label{fig:dacOut}
    \end{subfigure}
    \begin{subfigure}[b]{0.49\linewidth}
        \centering
        \includegraphics[width=\linewidth]{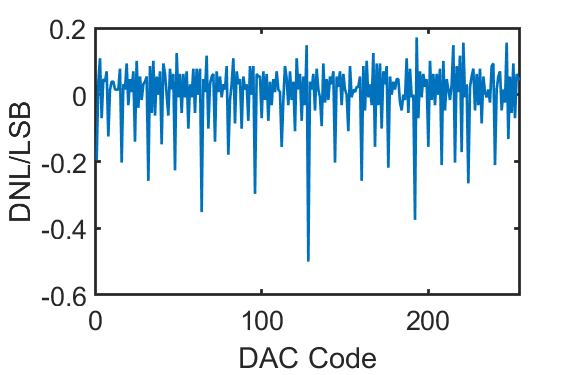}
        \caption{}
        \label{fig:dacDNL}
    \end{subfigure}
    \caption{Results from an automated test of an AD558J DAC (a) in $\SI{2.56}{\volt}$ range. (b) shows the output from code $0$ to code $255$. (c) shows the normalised differential non-linearity.}
\end{figure}

For our final test we exchanged the daughterboard used for previous test with a specialised version carrying a 48-pin ZIF socket (see Fig. \ref{fig:dacPic}) and used that to test an AD558J DAC \cite{Feedback2017}. We measured the input/output transfer characteristics (digital code to analogue output) and differential non-linearity (DNL). The IC was mounted in the ZIF socket (Fig. \ref{fig:dacPic}) and the board configured with jumpers to connect pins 1-8 to digital outputs and pin 11 to an onboard power supply. 

Pins 9, 10, 12, 13, and 16 were connected to analogue channels and pins 14 and 15 were shorted to pin 16 with jumper wires. The analogue channels on pins 9, 10, 12, and 13 were grounded and the supply pins was set to $\SI{10}{\volt}$. This configuration sets the IC as a $0-\SI{2.56}{\volt}$ DAC with transparent input latches. The digital inputs were then stepped through all input codes and the voltage measured at pin 16 at each step. We observed a maximum DNL of 0.5 LSB, matching the DNL specified on the datasheet. The daughterboard used here was configured with jumpers, but a version designed around analogue matrix switches could allow for a greater degree of automation.

\addtolength{\textheight}{-8.8cm}

\section{DISCUSSION AND CONCLUSIONS}\label{sec:disc}

In this paper, we have presented a general-purpose instrument that can accommodate the testing needs of a large variety of electronics component, featuring an appropriately large number of semi-independent source-metre channels. This new tool addresses an important demand for testing increasingly complex circuits while minimising the occasions where an entire PCB-based system needs to be designed to meet the requirements of the device under test. We note that important enabling factors for building such instrumentation include: a) the availability of FPGAs with large numbers of pins (allowing high parallelism), b) increasingly accessible multi-layer PCBs, c) the improvement of discrete components such as amplifiers and power supplies.

Importantly, the presented instrument achieves competitive specifications to several established benchtop instruments whilst remaining in a smaller, desktop format. Table \ref{tab:specs} summarises the achieved key performance metrics and compares them to other, established instruments in the area. Our results demonstrate how parallelism and portability can been traded against accuracy but not necessarily speed. While the low sampling rate of this system limits it to DC characterisation, the parallel structure allows for 1024 device arrays to be read in under $\SI{50}{\milli\second}$, facilitating the high throughput testing that is required by emerging technologies. Despite trading away some accuracy we are still at the point where the instrument can measure its own leakage currents, as well as set and measure all necessary noise floors (see Fig. \ref{fig:noise}). Furthermore, as demonstrated by the example applications the achieved accuracy is more than sufficient for supporting the needs of a very wide variety of electronic technologies. We thus foresee that this new tool will significantly aid the development as well as the use of emerging electron device technologies into new applications where read/write parallelism and data-level speed can be of paramount importance. We finally, acknowledge that the portability of the proposed instrument can be advantageous for a lab-at-home experience, particularly in light of the global pandemic ongoing at the time of writing this article.

In conclusion, we envisage that in the future versatile, portable instrumentation that can handle increasingly complex, non-accuracy-critical circuits will become more commonplace, accelerating and democratising research into electronic devices, components, chips etc. much the the Raspberry Pi and Arduino systems have done for embedded software research. We hope this new instrument will play a significant role in enabling this vision, as well as serve as a concrete example of such systems can be developed and what capabilities they can achieve.

\begin{table}[h]
    \centering
    \resizebox{\linewidth}{!}{    
        \begin{tabular}{|c||c|c|c|c|}
        \hline
             & \cite{DIGILENT2015} & \cite{ArcInstruments2017}  & \cite{Park2009} &  This work \\
        \hhline{|=||=|=|=|=|}
        Parallel read & N & N & N & Y\\
        \hline
        Parallel write & N & N & N & Y\\
        \hline
        Channel count & 2R+2W+16D & 32R+32W & 4R+2W & 64R/W+64D\\
        \hline
        Form factor & Portable & Desktop & Benchtop & Desktop\\
        \hline
        Min. chan. current & N/A & $\pm\SI{1}{\nano\ampere}$ & $\pm\SI{10}{\nano\ampere}$ & $\pm\SI{100}{\pico\ampere}$\\
        \hline
        Max. chan. current & N/A & $\pm\SI{5}{\milli\ampere}$ & $\pm\SI{500}{\milli\ampere}$ & $\pm\SI{12}{\milli\ampere}$\\
        \hline
        Current sample rate & N/A & $50-\SI{1000}{\siemens\per\second}$ & N/A & $\SI{833}{\siemens\per\second}$\\
        \hline
        Voltage resolution & $\SI{166/665}{\micro\volt}$ & $\SI{3/24}{\milli\volt}$ &$\SI{1}{\micro\volt}$& $\SI{78}{\micro\volt}$\\
        \hline
        Voltage sample rate & $\SI{100}{\mega\siemens\per\second}$ & $\SI{200}{\kilo\siemens\per\second}$ & $\SI{1.25}{\giga\siemens\per\second}$ & $\SI{100}{\kilo\siemens\per\second}$\\
        \hline
        Min. pulse width & N/A & $\SI{90}{\nano\second}$ & $\SI{10}{\nano\second}$ & $\SI{40}{\nano\second}$\\
        Max. chan. current & N/A & $\pm\SI{5}{\milli\ampere}$ & $\pm\SI{500}{\milli\ampere}$ & $\pm\SI{12}{\milli\ampere}$\\
        Pulse volt. range & $\pm\SI{5}{\volt}$ & $\pm\SI{12}{\volt}$ & $\pm\SI{20}{\volt}$ & $\pm\SI{13.5}{\volt}$\\
        \hline
        Power & $\SI{500}{\milli\watt}$ & $\SI{4.5}{\watt}$ & $\SI{2}{\watt}$ & $\SI{20}{\watt}$\\
        \hline
        \end{tabular}
    }
    \caption{Comparison between this work and similar systems.}
    \label{tab:specs}
\end{table}

\bibliography{ALL_library_3}

\begin{thebibliography}{10}
\providecommand{\url}[1]{#1}
\csname url@samestyle\endcsname
\providecommand{\newblock}{\relax}
\providecommand{\bibinfo}[2]{#2}
\providecommand{\BIBentrySTDinterwordspacing}{\spaceskip=0pt\relax}
\providecommand{\BIBentryALTinterwordstretchfactor}{4}
\providecommand{\BIBentryALTinterwordspacing}{\spaceskip=\fontdimen2\font plus
\BIBentryALTinterwordstretchfactor\fontdimen3\font minus
  \fontdimen4\font\relax}
\providecommand{\BIBforeignlanguage}[2]{{%
\expandafter\ifx\csname l@#1\endcsname\relax
\typeout{** WARNING: IEEEtran.bst: No hyphenation pattern has been}%
\typeout{** loaded for the language `#1'. Using the pattern for}%
\typeout{** the default language instead.}%
\else
\language=\csname l@#1\endcsname
\fi
#2}}
\providecommand{\BIBdecl}{\relax}
\BIBdecl

\bibitem{Input1998}
{Burr-Brown Corporation}, \emph{{INA111 Datasheet}}, 1998.

\bibitem{Feedback2017}
{Analog Devices}, \emph{{AD558 Datasheet}}, 2017.

\bibitem{2009char}
W.~{Chang}, K.~{See}, and B.~{Hu}, ``Characterization of component under dc
  biasing condition using an inductive coupling approach,'' \emph{IEEE
  Transactions on Instrumentation and Measurement}, vol.~59, no.~8, pp.
  2109--2114, 2010, DOI: 10.1109/TIM.2009.2031850.

\bibitem{pc-based}
M.~{Upadhye} and A.~P. {Sathe}, ``An integrated pc-based system for measurement
  of various parameters for two-terminal devices used in the industry,''
  \emph{IEEE Transactions on Instrumentation and Measurement}, vol.~41, no.~5,
  pp. 706--709, 1992, DOI: 10.1109/19.177347.

\bibitem{lock-in}
{Stanford Research System}, \emph{{SRA856A Datasheet}}.

\bibitem{Oscilloscopes}
Tektronix, \emph{{Tektronix MDO34 Datasheet}}.

\bibitem{Strukov08}
D.~B. Strukov \emph{et~al.}, ``{The missing memristor found},'' \emph{Nature},
  vol. 453, no. 7191, pp. 80--83, may 2008, DOI: 10.1038/nature06932.

\bibitem{Stathopoulos2019a}
S.~Stathopoulos \emph{et~al.}, ``{An Electrical Characterisation Methodology
  for Benchmarking Memristive Device Technologies},'' \emph{Scientific
  Reports}, vol.~9, no.~1, pp. 1--10, 2019, DOI: 10.1038/s41598-019-55322-4.

\bibitem{Serb2017HardwarelevelBI}
A.~Serb \emph{et~al.}, ``Hardware-level bayesian inference,'' in \emph{Neural
  Information Processing Systems}, December 2017.

\bibitem{Berdan2015b}
R.~Berdan \emph{et~al.}, ``{A $\mu$-Controller-Based System for Interfacing
  Selectorless RRAM Crossbar Arrays},'' \emph{IEEE Transactions on Electron
  Devices}, vol.~62, no.~7, 2015, DOI: 10.1109/TED.2015.2433676.

\bibitem{Xing2016}
J.~Xing \emph{et~al.}, ``{An FPGA-based instrument for en-masse RRAM
  characterization with ns pulsing resolution},'' \emph{IEEE Transactions on
  Circuits and Systems I: Regular Papers}, no.~6, pp. 818--826, jun 2016, DOI:
  10.1109/ISCAS.2016.7538870.

\bibitem{Knowm}
A.~Nugent, \emph{{Knowm Memristor Discovery manual}}, 2019.

\bibitem{Linn2010}
E.~Linn \emph{et~al.}, ``{Complementary resistive switches for passive
  nanocrossbar memories.}'' \emph{Nature materials}, vol.~9, no.~5, pp.
  403--406, may 2010, DOI: 10.1038/nmat2748.

\bibitem{Serb2015}
A.~Serb \emph{et~al.}, ``{Practical Determination of Individual Element
  Resistive States in Selectorless RRAM Arrays},'' \emph{IEEE Transactions on
  Circuits and Systems I: Regular Papers}, vol.~63, no.~6, pp. 827--835, jun
  2016.

\bibitem{Chen2013}
A.~Chen, ``{A Comprehensive Crossbar Array Model With Solutions for Line
  Resistance and Nonlinear Device Characteristics},'' \emph{IEEE Transactions
  on Electron Devices}, vol.~60, no.~4, pp. 1318--1326, apr 2013, DOI:
  10.1109/TED.2013.2246791.

\bibitem{Sivan2019}
M.~Sivan \emph{et~al.}, ``{All WSe2 1T1R resistive RAM cell for future
  monolithic 3D embedded memory integration},'' \emph{Nature Communications},
  vol.~10, no.~1, p. 5201, dec 2019, DOI: 10.1038/s41467-019-13176-4.

\bibitem{Stathopoulos2017}
S.~Stathopoulos \emph{et~al.}, ``{Multibit memory operation of metal-oxide
  bi-layer memristors},'' \emph{Scientific Reports}, no.~1, p. 17532, dec 2017.

\bibitem{opa3s328}
{Texas Instruments}, \emph{{OPA3S328 Datasheet}}, 2020.

\bibitem{Semiconductors2020}
N.~X.~P. Semiconductors, \emph{{LPC1769 Datasheet}}, 2020.

\bibitem{Wust2017}
D.~Wust \emph{et~al.}, ``Prototyping memristors in digital system with an
  fpga-based testing environment,'' in \emph{2017 27th International Symposium
  on Power and Timing Modeling, Optimization and Simulation (PATMOS)}, 2017,
  pp. 1--7, DOI: 10.1109/PATMOS.2017.8106978.

\bibitem{Wang2019}
Y.~Wang \emph{et~al.}, ``High speed test system of current pulse for phase
  change memory devices,'' \emph{Journal of Physics: Conference Series}, vol.
  1237, p. 042064, 06 2019, DOI: 10.1063/1.5042281.

\bibitem{Emmanuelle2016}
E.~Merced-Grafals \emph{et~al.}, ``Repeatable, accurate, and high speed
  multi-level programming of memristor 1t1r arrays for power efficient analog
  computing applications,'' \emph{Nanotechnology}, vol.~27, p. 365202, 08 2016,
  DOI: 10.1088/0957-4484/27/36/365202.

\bibitem{Jouppi2011}
N.~P. Jouppi, ``{Design implications of memristor-based RRAM cross-point
  structures},'' in \emph{2011 Design, Automation {\&} Test in Europe}.\hskip
  1em plus 0.5em minus 0.4em\relax IEEE, mar 2011, pp. 1--6, DOI:
  10.1109/DATE.2011.5763125.

\bibitem{Limited2021}
A.~R.~M. Limited, \emph{{AMBA AXI and ACE Protocol Specification}}, 2021.

\bibitem{cesys}
CESYS, \emph{{AXI-FX3-Interface v1.2}}, 2017.

\bibitem{Pattnayak}
Cypress, \emph{{EZ-USB FX3 SuperSpeed USB Controller}}, 2018.

\bibitem{DIGILENT2015}
DIGILENT, \emph{{Analog Discovery 2™ Reference manual}}, 2015.

\bibitem{ArcInstruments2017}
{Arc Instruments}, \emph{{Memristor Characterisation Platform User manual}},
  2017.

\bibitem{Park2009}
Keithley, \emph{{Semiconductor Characterization System Technical Data, Keithley
  4200-SCS.}}, 2009.

\end{thebibliography}


\begin{thebibliography}{99}

\bibitem{memristor} D. B. Strukov, G. S. Snider, D. R. Stewart, and R. S. William, “The missing memristor found,” Nature, vol. 453, no. 7191, pp. 80–83, 2008.

\bibitem{low-power} X. Yang and I.-W. Chen, “Dynamic-load-enabled ultra-low power multiple-state RRAM devices,” Sci. Rep., vol. 2, Oct. 2012, Art. ID 744.

\bibitem{brain-inspired} Xia, Qiangfei \& Yang, Jianhua Joshua. (2019). Memristive crossbar arrays for brain-inspired computing. Nature Materials. 18. 309-323.10.1038/s41563-019-0291-x.

\bibitem{1d1r} R. Aluguri and T. Tseng, "Notice of Violation of IEEE Publication Principles: Overview of Selector Devices for 3-D Stackable Cross Point RRAM Arrays," in IEEE Journal of the Electron Devices Society, vol. 4, no. 5, pp. 294-306, Sept. 2016. doi: 10.1109/JEDS.2016.2594190

\bibitem{1t1r} F. Bedeschi et al., “A bipolar-selecter phase change memory featuring multi-level cell storage,” IEEE Journal of Solid-State Circuits, vol. 44, no. 1, pp. 217–227, 2009.

\bibitem{dotproduct} Hu, Miao \& Graves, Catherine \& Li, Can \& Li, Yunning \& Ge, Ning \& Montgomery, Eric \& Dávila, Noraica \& Jiang, Hao \& Williams, Stan \& Yang, Jianhua Joshua \& Xia, Qiangfei \& Strachan, John William. (2018). Memristor‐Based Analog Computation and Neural Network Classification with a Dot Product Engine. Advanced Materials. 30. 10.1002/adma.201705914.

\bibitem{arcone} R. Berdan, A. Serb, A. Khiat, A. Regoutz, C. Papavassiliou and T. Prodromakis, "A $\mu $ -Controller-Based System for Interfacing Selectorless RRAM Crossbar Arrays," in IEEE Transactions on Electron Devices, vol. 62, no. 7, pp. 2190-2196, July 2015.

\bibitem{knowm} Knowm Inc. \emph{Memristor Discovery V2 Kit} \url{https://knowm.org/product/memristor-discovery-kit/}

\bibitem{Keithley} Tektronix inc. \emph{4200A-SCS Parameter Analyzer Datasheet} datasheet. 2019

\bibitem{Keysight} Keysight Technology.\emph{B1500A Semiconductor Device Analyzer} datasheet. 2019

\bibitem{bayesian}  Serb, Alexantrou, Manino, Edoardo, Messaris, Ioannis, Tran-Thanh, Long and Prodromakis, Themis "Hardware-level Bayesian inference". Neural Information Processing Systems. 7 pp . 2017
\end{thebibliography}

\begin{comment}

\end{comment}

\end{document}